\documentclass[conference]{IEEEtran}

\usepackage{balance}
\usepackage{algorithm,algorithmic}
\usepackage{graphicx}
\usepackage{multirow}
\usepackage{array}
\usepackage{ctable}
\usepackage[size=small,format=plain,labelfont=up,textfont=up]{caption}
\DeclareCaptionType{copyrightbox}
\usepackage{caption}
\usepackage{subcaption}
\usepackage{filecontents}
\usepackage{amsmath,amssymb}
\usepackage{epstopdf}
\usepackage{mathtools}

\newcommand{\tn}{\tabularnewline}

\renewcommand{\footnotesize}{\scriptsize}
\newcommand{\specialcell}[2][c]{%
  \begin{tabular}[#1]{@{}c@{}}#2\end{tabular}}

\newfont{\mycrnotice}{ptmr8t at 7pt}
\newfont{\myconfname}{ptmri8t at 7pt}
\let\crnotice\mycrnotice%
\let\confname\myconfname%

\begin{document}


\title{On Identifying Disaster-Related Tweets: Matching-based or Learning-based?}

\author{
\IEEEauthorblockN{Hien To, Sumeet Agrawal, Seon Ho Kim, Cyrus Shahabi}
\IEEEauthorblockA{Integrated Media Systems Center, University of Southern California, Los Angeles, USA\\
\{hto,sumeetag,seonkim,shahabi\}@usc.edu}
}

\maketitle

\begin{abstract}

Social media such as tweets are emerging as platforms contributing to situational awareness during disasters. Information shared on Twitter by both affected population (e.g., requesting assistance, warning) and those outside the impact zone (e.g., providing assistance) would help first responders, decision makers, and the public to understand the situation first-hand. Effective use of such information requires timely selection and analysis of tweets that are relevant to a particular disaster. Even though abundant tweets are promising as a data source, it is challenging to automatically identify relevant messages since tweet are short and unstructured, resulting to unsatisfactory classification performance of conventional learning-based approaches. Thus, we propose a simple yet effective algorithm to identify relevant messages based on matching keywords and hashtags, and provide a comparison between matching-based and learning-based approaches. To evaluate the two approaches, we put them into a framework specifically proposed for analyzing disaster-related tweets. Analysis results on eleven datasets with various disaster types show that our technique provides relevant tweets of higher quality and more interpretable results of sentiment analysis tasks when compared to learning approach.

\end{abstract}

\section{Introduction}
\label{sec:intro}

Enhancing situational awareness is of great importance for disaster response and recovery. Information shared on social media such as Twitter greatly enhance time-critical situational awareness~\cite{vieweg2012situational,fema2012crisis} by not only spreading the news about casualties and damages, rescue efforts and alerts but also providing on-topic information for disaster-affected population and first responders who may benefit from such information.
Twitter has been one of the most popular means of communication during disasters.
Particularly, geotagged tweets have been used to understand the situation in the affected areas, e.g., analyzing the speed and impact of an earthquake~\cite{sakaki2010earthquake} as shown in the news about the Napa Earthquake 2014, \emph{``Six Critically Injured, 120 Treated At Napa Hospital Following 6.0 Earthquake"}. 


Understanding tweet messages is challenging since they are short (maximum 144 characters) and informal, especially in disasters when identifying timely relevant information is critical. Thus, there has been a large body of work aiming to identify disaster-related tweets. Existing studies use either learning-based or matching-based approaches to select tweets that are relevant to a particular disaster. Learning-based approach builds a model from a set of labeled tweets and uses the model to predict another set of data (e.g.,~\cite{zhang2016semi,nguyen2016applications,olteanu2015expect,morstatter2013sample}). One challenge of learning-based approaches is that the accuracy of the trained model highly depends on the quality and size of training dataset. However, training datasets in the existing studies are often small because having large labeled data for training was demanding. Conventional matching-based approach enumerates a set of keywords and hashtags that are relevant to a particular disaster, and searches for the tweets containing those words (e.g.,~\cite{vu2016geosocialbound,cobo2015identifying,imran2014aidr,olteanu2014crisislex,ashktorab2014tweedr}). 
One drawback of this approach is that the set of manually defined keywords and hashtags may not be all-inclusive, i.e., users often use different unique hashtags for the same event.

In this study, we provide a comparative evaluation of both learning-based and matching-based approach in identifying disaster-related tweets. Furthermore, we propose an improved matching-based technique to better identify tweets that are relevant to a particular disaster type such as earthquake, flood. Our technique enhances conventional matching-based approach by effectively enlarging the number of relevant tweets and improving their quality. The technique is twofold. First, it searches for candidate hashtags in a collection of hashtags by matching a small set of core keywords. The core keywords are predefined for each disaster type while the hashtags can be extracted from a collection of tweets (i.e., tweet corpus). Thereafter, the candidate hashtags are refined by ruling out irrelevant ones through crowdsourcing. Both the refined hashtags and the core keywords are used to match relevant tweets.
To evaluate our matching-based technique, we use a complementary state-of-art technique that identifies relevant tweets by learning. Our technique applies a set of standard models including \emph{word2vec} for representing each tweet as an embedding vector, TF-IDF for penalizing high-frequency words, latent semantic indexing for dimension reduction, and logistic regression for classifying tweets into relevant and irrelevant ones. In order to have larger training datasets, we aggregate labeled Twitter data from multiple sources and group the tweets of the same disaster type.

Our experiments evaluate the two techniques by putting them into a proposed framework for analyzing geotagged tweets posted by the affected and unaffected population in disasters. Since there is no real ground truth in aggregated large datasets, the set of tweets selected by both methods is considered as a ground truth, from which their recall scores (the fraction of identified tweets that are relevant) are calculated as evaluation measurements. Experimental results on eleven disasters of three different types (earthquake, flood, wildfire) show that the matching-based technique provides a smaller number of relevant tweets but with higher quality (measured by the recall score) when compared to the learning-based technique. Our contributions are as follows.

\vspace{0.05in}
\noindent \textbf{1)} We identify the specific challenges of the two techniques in classifying relevant tweets: a large number of unique hashtags used for a particular disaster and a lack of big labeled datasets for training classification models.

\noindent \textbf{2)} We propose an improved matching-based technique that significantly increases the number of relevant tweets found when compared to traditional matching approaches (by up to 80\%).

\noindent \textbf{3)} We create eleven new datasets of geotagged tweets, each corresponds to a disaster occurred in 2014-2015. Our datasets (refined hashtags and tweets in affected/unaffected areas) are published with source code on GitHub\footnote{https://github.com/infolab-usc/bdr-tweet}, and open to other researchers.

\noindent \textbf{4)} We evaluate the matching-based technique on the datasets, which produces a set of relevant tweets with a higher quality when compared to the learning-based approach. Consequently, in a particular application of sentiment analysis, the matching-based technique also yields more interpretable results.

The remainder of this paper is organised as follows. Section~\ref{sec:related} discusses the related work, followed by our framework in Section~\ref{sec:framework}. Experimental results are presented in Section~\ref{sec:results}. Finally, we conclude the paper and discuss the future work in Section~\ref{sec:conclude}.

\section{Related Work}
\label{sec:related}

Social media such as Twitter and Facebook has been widely regarded as active communication channels during emergency events such as disasters caused by natural hazards. The Federal Emergency Management Agency (FEMA) identifies social media as an essential component of future disaster management~\cite{fema2012crisis}. Tweets sent during catastrophic events have been known to contain information that contributes to situational awareness~\cite{vieweg2012situational}, and a recent survey of studies for analyzing social media in disaster response can be found in~\cite{imran2015processing}.
Therefore, social media data analytics for disaster response has gained extensive interest from the research community.
Existing studies focus on extracting disaster-related information from socially-generated content during natural disasters, from which actionable information can be disseminated to disaster relief workers~\cite{imran2013practical}. More recent studies build classifiers for identifying earthquake-relevant tweets~\cite{cobo2015identifying}, classifying tweets based on informative and uninformative tweets~\cite{parilla2014disaster,nguyen2016applications,zhang2016semi,caragea2016identifying}.
Furthermore, tweets can be categorized by type~\cite{vieweg2010microblogging,imran2013extracting,caragea2011classifying,acar2011twitter} (i.e., affected individuals, infrastructure and utilities, donations and volunteer, caution and advice, sympathy and condolence) or by information source~\cite{starbird2011voluntweeters,monroy2013new,diakopoulos2012finding,de2012unfolding,vieweg2010microblogging} (i.e., eyewitness, government, NGOs, business, media).

Analyzing public sentiment during disasters is a popular application of social media for disaster response. Although the problem has been extensively studied in other domains, such as product reviews~\cite{pang2008opinion,liu2012sentiment}, understanding public sentiment from social media is gradually applied in disasters. However, this problem is challenging as the social media content are often short and unstructured~\cite{hu2013unsupervised,hu2013exploiting}.  There has been a growing body of work addressing such challenges~\cite{caragea2014mapping,lu2015visualizing}. In~\cite{caragea2014mapping}, sentiment classification of twitter messages during Hurricane Sandy was performed and the extracted sentiments were visualized on a geographical map centered around the hurricane. In~\cite{lu2015visualizing}, an entropy-based metric was proposed to model sentiment contained in tweet messages. The extracted sentiments were visualized through a map-based interface to reveal interesting patterns in disasters.

These studies typically used machine learning tools to filter disaster-related tweets. An issue with such approaches is the lack of appropriate interpretation of the results, e.g., why a technique works well on some data (high precision and recall) but do not perform as well in others~\cite{imran2014aidr,ashktorab2014tweedr}.
In addition, the studies tend to examine one particular disaster and imply that the findings are generalizable to others ~\cite{fraustino2012social}. However, it is known that information shared on Twitter varies considerably from one crisis to another~\cite{kanhabua2013understanding,olteanu2014crisislex,olteanu2015expect}. We aim to narrow the gap by customizing each component of our framework for a particular disaster.







\section{Framework}
\label{sec:framework}

\begin{figure}[ht]
		\centering
		\includegraphics[width=0.49\textwidth]{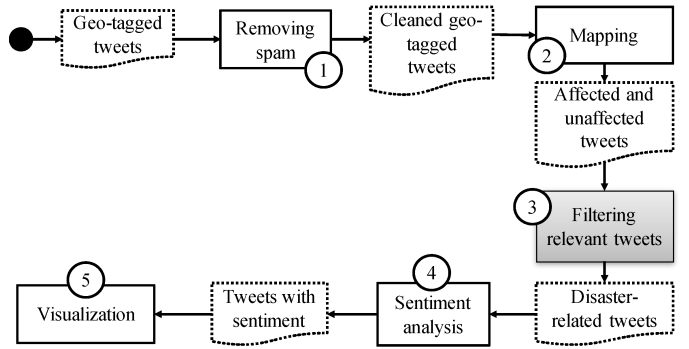}
		\caption{A framework for analyzing geo-tagged tweets generated during disasters. The focus of this study is highlighted as shaded boxes.}
		\label{fig:framework}
\end{figure}

We propose a five-step framework for processing and analyzing disaster-related tweets as shown in Figure~\ref{fig:framework}. Note that we consider only geo-tagged tweets in this study. For each disaster, spam tweets are removed (step 1). Thereafter, the cleaned geo-tagged tweets are mapped to affected and unaffected regions in the vicinity of each disaster (step 2). For each region, tweets are categorized into relevant and irrelevant ones (step 3), in which relevant tweets are analyzed to identify the popularity of sentiments expressed by users during the disasters (step 4). Finally, spatial and temporal patterns and trends of the mapped sentiments are revealed through visualization (step 5). In the following we detail each phase of the framework.

\subsection{Removing spam tweets and mapping tweets into affected and unaffected areas}

Many tweets are generated by spammers or non-human bots. Those tweets, if large enough, may falsify the results of analysis. Therefore, it is important to remove those spam tweets at the first step. We eliminate tweets from a user who generates more than a certain number of tweets (in this study, 15 tweets per day). Note that we do not remove retweets (RTs) as we are interested in the actions of reposting or forwarding existing tweets.

In the second step, we categorise tweets based on geographical regions, i.e., regions affected by a disaster and outside of the impact zone. We are specifically interested in the messages from the affected population as they usually require more and timely attention or assistance. According to FEMA, there are two kinds of assistance: individual assistance (e.g., damage to impacted residences) and public assistance (e.g., repair or replacement of facilities). A declaration for individual or public assistance for counties is requested by the Governor. Affected regions are considered to be the counties that need assistance while unaffected regions are other nearby counties without any assistance. The result is a clean dataset from affected regions.

\subsection{Identifying relevant tweets}
\label{sec:relevant}
Typically, there are two approaches in identifying tweets that are relevant to a specific disaster: matching-based~\cite{cobo2015identifying,imran2014aidr,olteanu2014crisislex,ashktorab2014tweedr,imran2013extracting,imran2013practical} and learning-based~\cite{zhang2016semi,nguyen2016applications,olteanu2014crisislex,olteanu2015expect,morstatter2013sample,hu2013unsupervised,caragea2011classifying}. 

\subsubsection{Matching-based approach}

The studies in this group typically use a set of keywords or hashtags to determine relevant tweets by identifying them in messages. An issue with the existing studies in this approach is that they typically use a small set of predefined hashtags such as combining disaster name/type with the name of affected area (e.g., \emph{\#napaearthquake}) or official name of the disaster (e.g., \emph{\#hurricanesandy}). However, such a simple approach may miss many relevant hashtags generated and used by users, for example, \emph{\#3amearthquake, \#staysafenapa, \#fearoftheearthquake}, which are diverse in terms of word choices. Also, many such hashtags are misspelled, such as \emph{\#eathquake, \#eartquake, \#earrhquake}, which cannot be detected by the existing simple solution. The performance of the matching-based approach relies on the completeness of the used keywords and hashtags. Our approach is to systematically construct a complete set of keywords and hashtags.

\begin{figure}[ht]
	\centering
	\begin{minipage}[b]{0.9\linewidth}
		\includegraphics[width=1\textwidth]{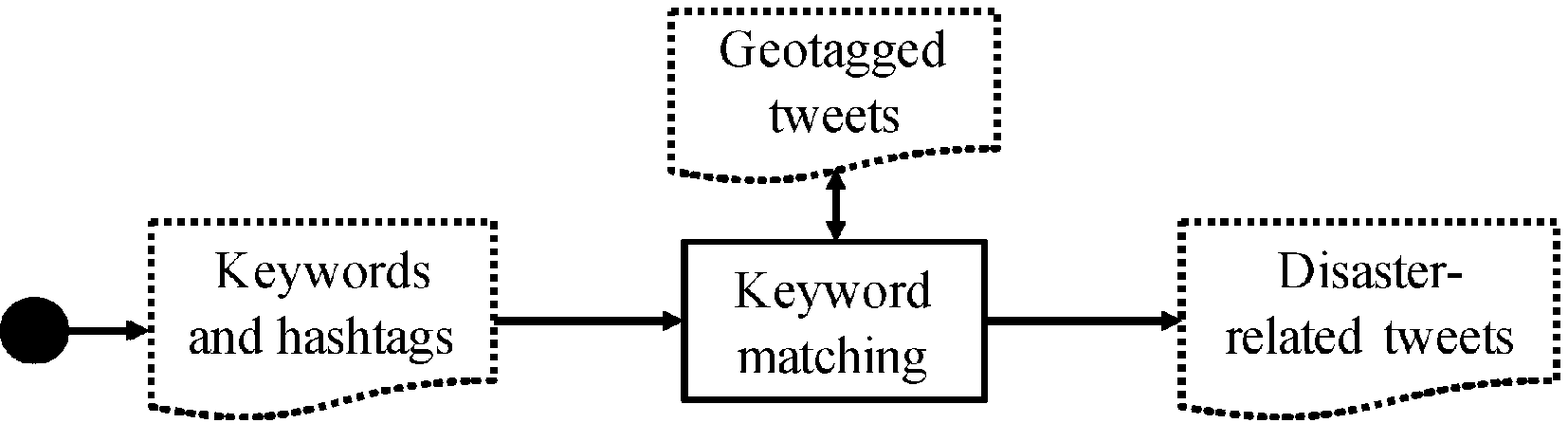}
		\subcaption{Conventional matching-based method}
		\label{fig:conventional_filter}
	\end{minipage}
	\centering
	\begin{minipage}[b]{0.9\linewidth}
	\vspace{10pt}	\includegraphics[width=1\textwidth]{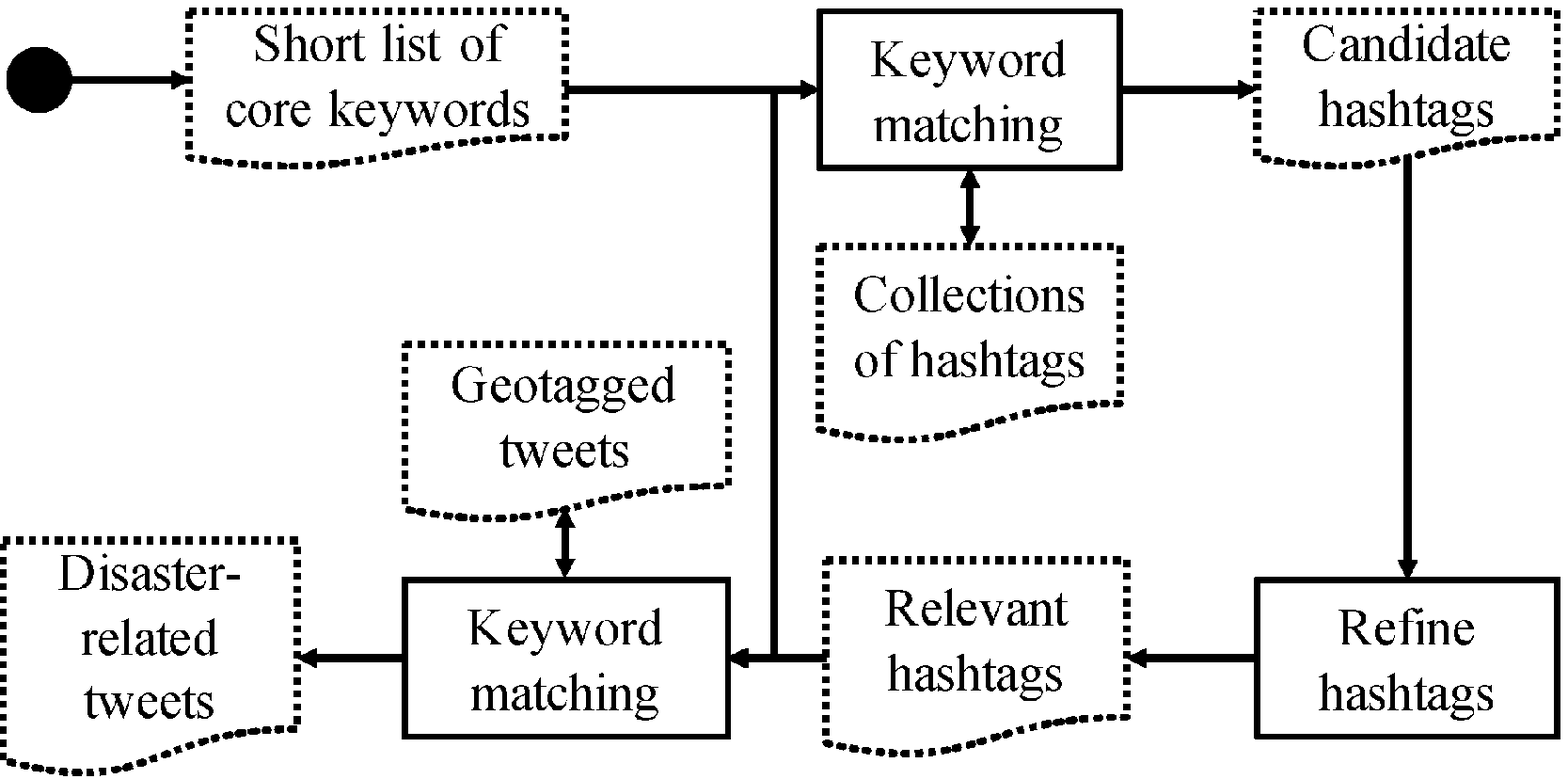}
		\subcaption{Matching-based method}
		\label{fig:matching_filter}
	\end{minipage}
	\centering
	\begin{minipage}[b]{0.9\linewidth}
	\vspace{10pt}	\includegraphics[width=1\textwidth]{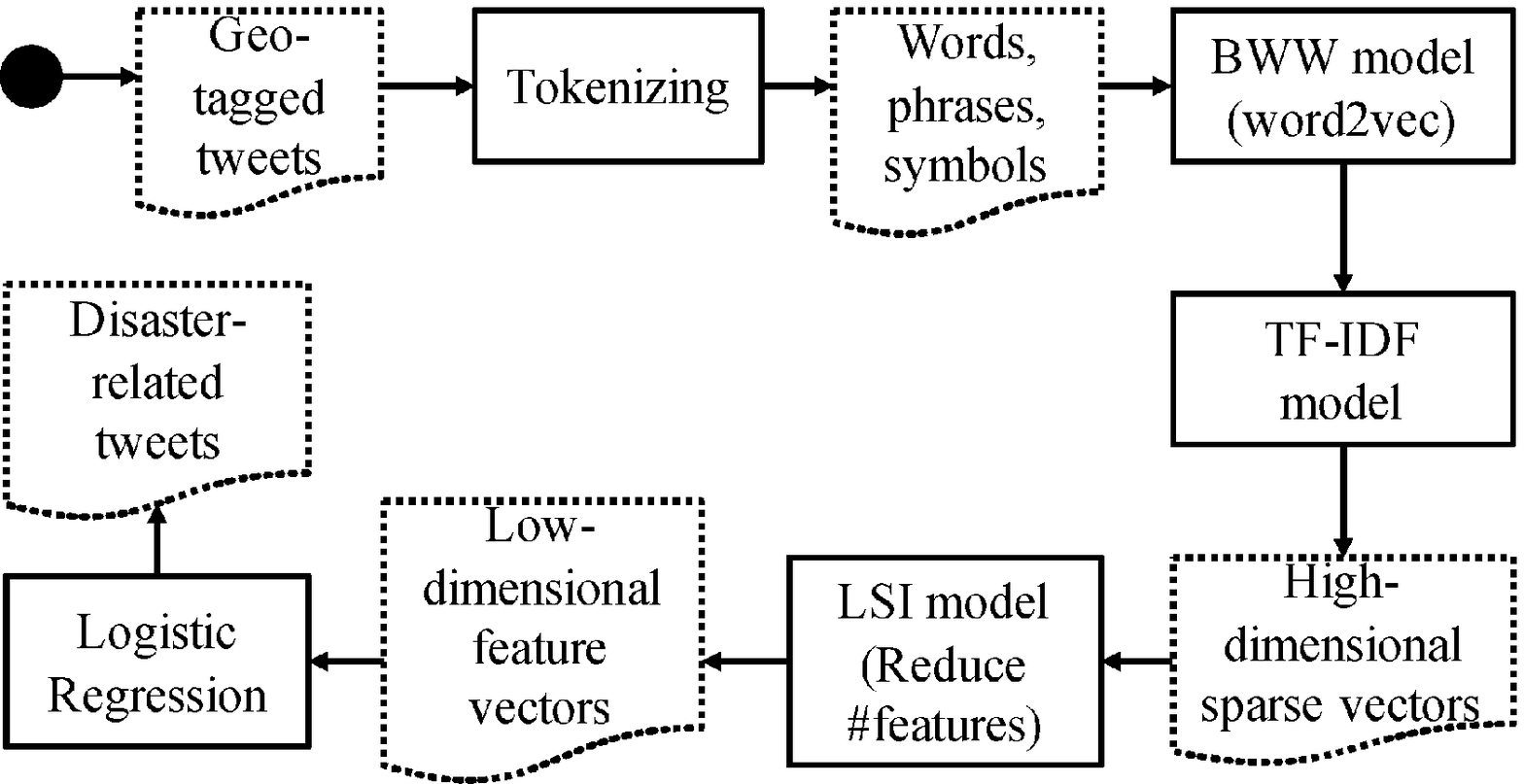}
		\subcaption{Learning-based method}
		\label{fig:learning_filter}
	\end{minipage}
	\caption{Methods to select disaster-relevant tweets.}
	\label{fig:filter_method}
    \vspace{-10pt}
\end{figure}

Observing that each disaster type has a small set of core keywords (see column 3 of Table~\ref{tab:hashtags}) like existing studies, we use these core keywords to search for more relevant hashtags on a dictionary of all hashtags, retrieved from the collection of tweets. For each disaster, the dictionary of hashtags can be retrieved by a linear scan through tweet collection to increase the number of disaster-related hashtags. However, simple word matching may include keywords irrelevant to disasters, for example, candidate hashtag \emph{``\#fireworks"} contains keyword \emph{``\#fire"} but it is unlikely related to wildfires. Such automatic semantic analysis is very challenging and still hard to achieve in practice. Hence, we refine all candidate hashtags to improve the quality of the hashtag collection through crowdsourcing. Crowd reviewers discard relevant hashtags, producing a set of related hashtags only. The number of candidate hashtags and refined hashtags for each disaster in our study is shown in the last two columns of Table~\ref{tab:hashtags}. On the other hand, the refined hashtags may not cover all the keywords corresponding to a particular disaster type (when the keywords occur in the tweet messages except the hashtags). Therefore, we combine the refined hashtags with the keywords, creating the final list of keywords and refined hashtags for each disaster. These terms are used to search tweets relevant to a given particular disaster. The matching-based approach is depicted in Figure~\ref{fig:matching_filter}.

\begin{table*}[h]
\begin{center}
\begin{tabular}{|l | *{4}{c|}}
\hline
{\bf DisasterId}	 & {\bf Type}	& {\bf Core Keywords} & {\bf Candidate Hashtags} & {\bf Refined Hashtags} \\
\hline
napa\_earthquake & earthquake & \parbox{1.5in}{quake, tremor, foreshock, aftershock} & 300  & 108
\tn\cline{1-5}
michigan\_storm & \multirow{9}{*}{flood} & \multirow{9}{*}{\parbox{1.5in}{flood, storm, typhoon, tornado, hurricane, mudslide, strong wind, high water}} & 89 & 87 \tn\cline{1-1}\cline{4-5}
newyork\_storm & \multicolumn{1}{c |}{} & \multicolumn{1}{c |}{} & 832 & 705 \tn\cline{1-1}\cline{4-5}
texas\_storm & \multicolumn{1}{c |}{} & \multicolumn{1}{c |}{}  & 166 & 125 \tn\cline{1-1}\cline{4-5}
iowa\_stf & \multicolumn{1}{c |}{} & \multicolumn{1}{c |}{} & 160 & 124 \tn\cline{1-1}\cline{4-5}
iowa\_stf2 & \multicolumn{1}{c |}{} & \multicolumn{1}{c |}{} & 168 & 116 \tn\cline{1-1}\cline{4-5}
iowa\_storm & \multicolumn{1}{c |}{} & \multicolumn{1}{c |}{} & 20 & 13 \tn\cline{1-1}\cline{4-5}
washington\_storm & \multicolumn{1}{c |}{} & \multicolumn{1}{c |}{} & 65 & 33 \tn\cline{1-1}\cline{4-5}
jersey\_storm & \multicolumn{1}{c |}{} & \multicolumn{1}{c |}{} & 20 & 18 \tn\cline{1-5}
california\_fire  & wildfire & \parbox{1.5in}{fire, firing, burn, buring, blaze, blazing, flame, framing} & 318 & 149 \tn\cline{1-5}
washington\_mudslide & flood + wildfire & \emph{keywords of both flood and wildfire} & 108 & 64 \\
\hline
\end{tabular}
\caption{Summary of used keywords and hashtags.}
\label{tab:hashtags}
\end{center}
\end{table*}

An advantage of our solution over conventional matching-based approaches is the ability to systematically generate a comprehensive list of relevant hashtags starting from a small set of core keywords. The number of relevant hashtags can be as large as a few hundred (see Table~\ref{tab:hashtags}) while existing studies~\cite{cobo2015identifying,imran2014aidr,ashktorab2014tweedr,imran2013extracting,imran2013practical} often use less than ten hashtags.
Nevertheless, there may be still missing relevant tweets that do not contain the keywords or hashtags but are semantically relevant to a disaster. An example of such type of tweet is ``\emph{@Securb Safe and Sound! Just a lot of shaking in SF but no damage! Thanks@for checking!}." This tweet is relevant to Napa earthquake although it does not contain any core keywords or refined hashtags of the disaster.


\subsubsection{Learning-based approach}
\label{sec:learning}
Learning-based approach tends to include more tweets into relevant sets when compared to the matching-based method. 
The accuracy of learning-based techniques often highly depends on the quantity and quality of training datasets. However, to best our knowledge there is no large labeled dataset available for classifying disaster-related tweets. Thus, we prepare our own training datasets from multiple data sources. In the following, we present our five-step solution (see Figure~\ref{fig:learning_filter}) to classify tweets into relevant and irrelevant ones: 1) tokenizing each tweet into words, phrases and symbols called tokens, 2) representing tweets using bag-of-words model, 3) penalizing high-frequency words using TF-IDF model, 4) projecting tweets to fixed-size low-dimensional feature vectors, and 5) using logistic regression for classification.

\paragraph{Preparing training datasets}
Existing studies~\cite{olteanu2014crisislex,olteanu2015expect} filter tweets relevant to a specific disaster by building a separate model (or classifier) from a specific training dataset for each disaster (e.g., one for Colorado floods and another for Alberta floods). A challenge of this approach is to have large labeled data for training the models. However, labeled datasets used in these studies are often small, e.g., each training dataset in~\cite{olteanu2014crisislex} has only about 1000 tweets.
 
To mitigate this issue, unlike~\cite{olteanu2014crisislex,olteanu2015expect}, we build the same model for all disasters of the same type. The reason for this is twofold. First, we improve the models by enlarging the sizes of the training datasets. Second, most tweets related to a disaster type share the same set of core keywords. Consequently, for every kind of the disaster (earthquake, flood, wildfire), we create different training data from publicly available data sources, including CrisisLexT26~\cite{olteanu2014crisislex} and CrowdFlower10K\footnote{http://www.crowdflower.com/data-for-everyone}. Note that we also train one hybrid model of flood and wildfire for \emph{washington\_mudslide}. A summary of the two data sources and the combined training datasets for each disaster type are shown in Table~\ref{tab:labeled_datasets}. CrisisLexT26 includes tweets collected during 26 large crisis events in 2012 and 2013, with about 1,000 labeled tweets per crisis. CrowdFlower10K contains 10,876 tweets relevant to various kinds of disaster, from major ones that cause damaged structures to minor ones such as car accidents. This dataset was created in two steps: 1) automatically searching for tweets with keywords such as \emph{``ablaze"} and \emph{``quarantine"}, and then 2) tweets are manually classified by CrowdFlower's workers into one of the three labels: \emph{``Relevant"}, \emph{``Not Relevant"}, and \emph{``Can't Decide"}. In order to enhance the quality of the training data, we discarded the tweets with \emph{``Can't Decide"} label and the ones with confidence scores less than one (i.e., only keep tweets with 100\% confidence). The confidence score is determined by CrowdFlower's workers. Finally, we use the keywords and hashtags in Table~\ref{tab:hashtags} to find tweets relevant to each disaster.

\begin{table*}[h]
\begin{center}
\begin{tabular}{| l | *{5}{c|}}
\hline
\multirow{2}{*}{{\bf Disaster type}} &\multicolumn{2}{ c |}{{\bf CrisisLexT26}} & \multicolumn{2}{ c |}{{\bf CrowdFlower10K}} & \multirow{2}{*}{{\bf Combined training datasets}} \tn\cline{2-5}
\multicolumn{1}{| c |}{} & Related & Not Related & Related & Not Related & \multicolumn{1}{c |}{}  \tn\cline{1-6}
Earthquake & 2,334 & 2,132 & 57 & 35 & 4,561 \tn\cline{1-6}
Flood  & 5,121 & 3,135 & 795 & 382 & 9,447 \tn\cline{1-6}
Wildfire & 2,015 & 1,387 & 122 & 33 & 3,557 \tn\cline{1-6}
\hline
\end{tabular}
\caption{Summary of labeled datasets for training.}
\label{tab:labeled_datasets}
\end{center}
\end{table*}

\paragraph{Tokenizing tweets}
For each tweet, we use our customized tokenizer to remove non-ASCII characters, HTML tags and replace emojis with corresponding words (e.g., \emph{``;)"} by \emph{``happy")}. We also attempt to split hashtags into meaningful elements if possible (e.g., \emph{``californiaearthquake"} to \emph{``california"} and \emph{``earthquake"}).

\paragraph{Representing tweets using bag-of-words model}

We use a bag-of-words (BOW) model to represent each tweet as a vector where each dimension denotes a particular word and a value which specifies the number word appearance in a particular tweet.
Specifically, we use a powerful word embedding technique named \emph{word2vec}~\cite{mikolov2013distributed}, which was also used in many recent studies~\cite{zhang2016semi,nguyen2016applications}, to create the BOW model. We use an implementation of \emph{word2vec} provided by Gensim library~\cite{rehurek_lrec}. The input of \emph{word2vec} is a tweet corpus. It outputs a set of high-dimensional \emph{sparse} vectors (one for each tweet) as most words in our collection of tweets do not appear in a particular tweet. In this study our tweet corpus contains every word that appears at least twice in the whole tweet corpus. With \emph{word2vec}, the vectors of similar words are grouped together in the vector space. As a result, the word vectors capture many linguistic regularities~\cite{mikolov2013distributed}. For example, the cosine similarity of \emph{vector(``earthquake")} and \emph{vector(``aftershock")} would be higher than that for \emph{vector(``earthquake")} and \emph{vector(``flood")}, or vector operations \emph{vector(``earthquake")-vector(``geophysical")+vector(``tornado")} would result in a vector that is very close to \emph{vector(``meteorological")}.

\paragraph{Penalizing high-frequency words using TF-IDF model}

We apply a well-known statistic in information retrieval and text mining, named term frequency-inverse document frequency (TF-IDF), to factor the importance of a word in the corpus. This model diminishes the weight of words that occur very frequently in the tweet corpus such as ``the" and increases the weight of terms that occur rarely such as ``earthquake". Toward that end, each word count in the BOW model is penalized by multiplying with an inverse function of the number of tweets containing the word.

\paragraph{Projecting tweets to fixed-size low-dimensional feature vectors}

Previous studies~\cite{cobo2015identifying,morstatter2013sample,hu2013unsupervised,caragea2011classifying} have used dimensionality reduction techniques to reduce the number of features of classifiers.
Similarly, we use a popular topic-modeling technique, termed Latent Semantic Indexing (LSI), to transform the high-dimensional BOW space to a lower dimensional latent space. LSI uses Principal Component Analysis (PCA), which is a popular approach to map high-dimensional data into low-dimensional latent space while preserving as much variance as in the high-dimensional data as possible with the rapidly diminishing return for each new dimension. The output of this phase is a set of fixed-size low-dimensional feature vectors, which is fed into the logistic regression classifier. We use a popular implementation of logistic regression supported by the \emph{scikit-learn} library in Python.

\subsection{Analyzing sentiment of relevant tweets}

In this section, we present sentiment analysis as a particular application of the disaster-related tweets identified by the prior phase (see Section~\ref{sec:relevant}). We use this application to compare the results of the two approaches for identifying relevant tweets. To identify sentiment of individual tweets, we adopt a popular word embedding technique, named \emph{doc2vec}, which has been shown to achieve state-of-the-art results for sentiment analysis tasks~\cite{le2014distributed}.
In~\cite{le2014distributed}, each document (or tweet in this case) is represented by a fixed-length vector. 

We use an implementation of the \emph{doc2vec} algorithm provided by the Gensim library\footnote{https://radimrehurek.com/gensim/models/doc2vec.html}.
We feed into the \emph{doc2vec} model three kinds of data 1) a publicly available set of 1.6 millions labeled tweets (Sentiment140\footnote{http://help.sentiment140.com/for-students/}) with an equal number of positive and negative ones for \emph{training}, 2) a small set of labeled tweets for \emph{testing}, and 3) our disaster-relevant tweets for \emph{prediction}. The model outputs three sets of corresponding vectors. We use the training vectors to train a scikit-learn logistic regression classifier and the testing vectors to evaluate the accuracy of the sentiment classifier. Thereafter, we use the classifier to predict labels for the prediction vectors. 

Using the Gensim tool, we trained the \emph{doc2vec} model using 2,450,000 tweets with word vector dimension of 100 and vocabulary size of 2,178,060. We run \emph{doc2vec} with a negative sampling rate of $10^{-4}$, context window size of 10, and we do not ignore words with low total frequency.


\section{Results}
\label{sec:results}

\subsection{Experimental Setup}
\label{sec:setup}

We used a publicly available dataset in~\cite{pfeffer2016geotagged}, which consists of IDs of geotagged tweets within the U.S during two time periods: June to November in 2014 and 2015. From the Tweet IDs, we created eleven datasets using Twitter streaming API, each corresponds to a particular disaster that was officially declared by FEMA during the time periods. We considered three specific types of disasters in our experiments: flood, earthquake and wildfire. Flood is the most frequent type of disaster in our datasets, and also the most common natural hazard in the world. 
A summary of these datasets is given in Table~\ref{tab:disasters}, including various statistics such as the number of counties in the vicinity of the disasters and the number of affected counties. This data can be obtained from FEMA website, from which we computed the total number of geotagged tweets in the vicinity area and the number of those in the affected area. By default, we used the disaster-related tweets in the affected counties for evaluation.


\begin{table*}[h]
\begin{center}
\begin{tabular}{| l | *{7}{c|}}
\hline
{\bf DisasterId}	 & {\bf FEMA Code} & {\bf Start Date}	&	{\bf Duration (days)} & {\bf Tweets} & {\bf Counties} & {\bf Affected Tweets} & {\bf Affected Counties}  \\
\hline
napa\_earthquake & 4193 & 08-24-2014 & 16 & 1,868,964  & 58 & 374,782  & 2 \\
\hline
michigan\_storm & 4195 & 08-11-2014 & 3 & 399,293 & 83 & 1,90,394 & 3 \\
\hline
newyork\_storm & 4204 & 11-17-2014 & 11 & 227,073 & 62 & 143,505 & 9 \\
\hline
texas\_storm & 4245 & 10-22-2015 & 10 & 231,808 & 254 & 72,088 & 22 \\
\hline
iowa\_stf & 4184 & 06-14-2014 & 11 & 239,588 & 99 & 41,471 & 26 \\
\hline
iowa\_stf2 & 4187 & 06-26-2014 & 13 & 274,954 & 99 & 74,355 & 24 \\
\hline
iowa\_storm & 4234 & 06-20-2015 & 6 & 32,286 & 99 & 5,721 & 19 \\
\hline
washington\_storm & 4242 & 08-29-2015 & 1 & 79,381 & 39 & 9,217 & 6 \\
\hline
jersey\_storm & 4231 & 06-23-2015 & 1 & 114,925 & 21 & 20,406 & 4 \\
\hline
california\_fire & 4240 & 09-09-2015 & 52 & 430,253 & 58 & 1,916 & 2 \\
\hline
washington\_mudslide & 4,243 & 08-09-2015 & 33 & 80,188 & 39 & 6,260 & 8 \\
\hline
\end{tabular}
\caption{Statistics of 11 datasets.}
\label{tab:disasters}
\end{center}
\end{table*}


\subsection{Experimental Results}

\subsubsection{Comparison of the matching and learning approaches in identifying relevant tweets}

\begin{figure}[ht]
	\centering
	\begin{minipage}[b]{0.49\linewidth}
		\includegraphics[width=1\textwidth]{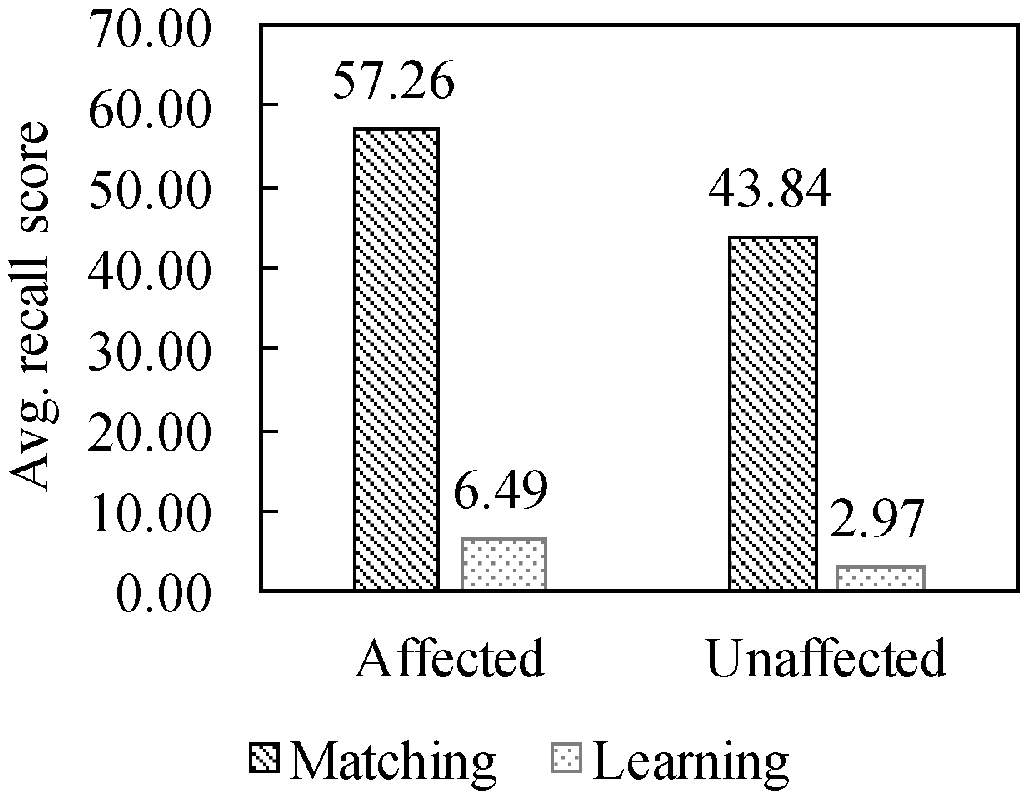}
		\subcaption{Accuracy of retrieved tweets}
		\label{fig:recall}
	\end{minipage}
	\centering
	\begin{minipage}[b]{0.49\linewidth}
		\includegraphics[width=1\textwidth]{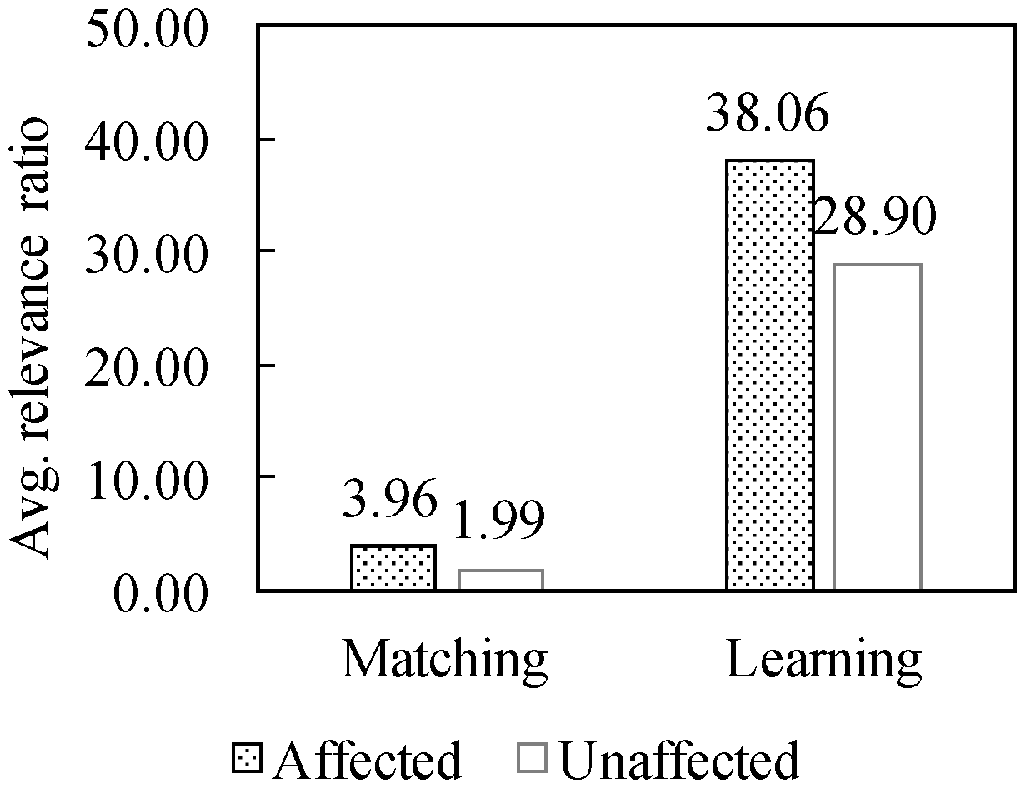}
		\subcaption{Percentage of related tweets}
		\label{fig:relevance}
	\end{minipage}
	\caption{Comparison of matching vs. learning-based approaches (average over 11 datasets).}
	\label{fig:relevance_recall}
\end{figure}

We compute the number of disaster relevant tweets that are selected using both classifiers described in Section~\ref{sec:relevant}. The results are shown in Table~\ref{tab:results}. We observe that the number of relevant tweets found by the learning-based classifier is much higher than that by the matching-based classifier. The reason for this might be twofold: the learning technique may include many irrelevant tweets and the matching technique would miss many relevant tweets. To elaborate this issue, as demonstrated in~\cite{lu2015visualizing,caragea2014mapping}, we assume that the set of agreed tweets by both classifiers has a higher accuracy  than the results from an individual classifier and thus can be considered as a ground truth. Given such assumption, the fraction of relevant tweets that are retrieved (precision score) is always 1. As another measurement, we compute the \emph{recall} of each classifier, which is the fraction of retrieved tweets that are relevant. The recall scores of each classifier in various disaster cases are shown in Table~\ref{tab:results}. We observe that the recall of the matching-based approach is an order of magnitude higher than that of the learning-based approach for all cases. This indicates that the matching-based technique outputs a smaller number of relevant tweets of higher quality when compared to the learning-based technique which tends to include more tweets with low quality. To summarize, we compute the average recall score of all disasters as presented in Figure~\ref{fig:recall}. We also observe that the recall scores for the affected regions are significantly higher than that in the unaffected areas. To explain this result, we show the number of disaster-related tweets found by our filtering techniques in both affected and unaffected regions. 

Particularly, we are interested in the percentage of related tweets retrieved, the number of relevant tweets divided by the total number of tweets, and refer to this value as \emph{relevance ratio}. Table~\ref{tab:results} shows the ratio for both affected and unaffected regions. As expected, the relevance ratio in the affected regions is significantly higher than that in the \emph{nearby} unaffected regions. This observation is true for all cases in our experiments showing that the population in affected areas are more likely to post disaster-related tweets. This result also explains the prior observation that the recall scores are higher in the affected areas when compared to the unaffected area.
We also show average relevance ratio over all disasters (see Figure~\ref{fig:relevance}), which shows that, with the matching-based approach, the number of relevant tweets found in the affected areas is twice of that in the unaffected areas.

\begin{table*}[h]
\begin{center}
\begin{tabular}{|l | *{11}{c|}}
\hline
\multirow{2}{*}{{\bf DisasterId}} & \multirow{2}{*}{\specialcell{\textbf{Spam}\\\textbf{Ratio (\%)}}} & \multirow{2}{*}{{\bf Area}}  & \multicolumn{3}{ c |}{{\bf Number of Disaster-Related Tweets}} & \multicolumn{2}{ c |}{{\bf Recall Score}}   & \multicolumn{2}{ c |}{{\bf Relevance Ratio}} \tn \cline{4-10}  
\multicolumn{1}{| c |}{} & \multicolumn{1}{c|}{} & \multicolumn{1}{c|}{} & Matching & Learning & Agreement & Matching & Learning  & Matching & Learning \tn\cline{1-10} 
\multirow{2}{*}{napa\_earthquake}  & \multirow{2}{*}{26.00} & Affected & 8,548 & 116,187 & 3,948 & 46.19 & 3.40 & 2.92 & 39.75 \tn 
\multicolumn{1}{| c |}{} & \multicolumn{1}{c |}{}  & Unaffected & 851 & 55,678 & 430 & 50.53 & 0.77 & 0.08 & 5.10 \tn\cline{1-10} 
\multirow{2}{*}{michigan\_storm} & \multirow{2}{*}{28.81} & Affected  & 2,638 & 31,129 & 1,183 & 44.84 & 3.80 & 2.07 & 24.40 \tn 
\multicolumn{1}{| c |}{} & \multicolumn{1}{c |}{}  & Unaffected & 1,767 & 38,811 & 689 & 38.99 & 1.78 & 1.13 & 24.77 \tn\cline{1-10} 
\multirow{2}{*}{newyork\_storm} & \multirow{2}{*}{24.69} & Affected & 6,952 & 29,412 & 3,786 & 54.46 & 12.87 & 6.73 & 28.47 \tn 
\multicolumn{1}{| c |}{} & \multicolumn{1}{c |}{}  & Unaffected & 1,611 & 19,154 & 793 & 49.22 & 4.14 & 2.38 & 28.30 \tn\cline{1-10} 
\multirow{2}{*}{texas\_storm} & \multirow{2}{*}{20.82}& Affected  & 2,871 & 37,044 & 2,237 & 77.92 & 6.04 & 4.74 & 61.18 \tn 
\multicolumn{1}{| c |}{} & \multicolumn{1}{c |}{}  & Unaffected & 4,251 & 37,921 & 1,561 & 36.72 & 4.12 & 3.46 & 30.83 \tn\cline{1-10} 
\multirow{2}{*}{iowa\_stf} & \multirow{2}{*}{19.61} & Affected & 1,756 & 8,031 & 933 & 53.13 & 11.62 & 4.87 & 22.26 \tn 
\multicolumn{1}{| c |}{} & \multicolumn{1}{c |}{}  & Unaffected & 3,782 & 37,304 & 1,702 & 45.00 & 4.56 & 2.42 & 23.83 \tn\cline{1-10} 
\multirow{2}{*}{iowa\_stf2} & \multirow{2}{*}{20.11} & Affected & 2,010 & 17,937 & 1,193 & 59.35 & 6.65 & 3.18 & 28.38 \tn 
\multicolumn{1}{| c |}{} & \multicolumn{1}{c |}{}  & Unaffected & 4,145 & 36,501 & 1,821 & 43.93 & 4.99 & 2.65 & 23.33 \tn\cline{1-10} 
\multirow{2}{*}{iowa\_storm} & \multirow{2}{*}{17.87} & Affected & 192 & 1,112 & 61 & 31.77 & 5.49 & 3.65 & 21.12 \tn 
\multicolumn{1}{| c |}{} & \multicolumn{1}{c |}{}  & Unaffected & 442 & 1,926 & 57 & 12.90 & 2.96 & 2.08 & 9.06 \tn\cline{1-10} 
\multirow{2}{*}{washington\_storm} & \multirow{2}{*}{13.95} & Affected & 283 & 4,657 & 179 & 63.25 & 3.84 & 3.27 & 53.75 \tn 
\multicolumn{1}{| c |}{} & \multicolumn{1}{c |}{}  & Unaffected & 1,873 & 26,976 & 980 & 52.32 & 3.63 & 3.14 & 45.23 \tn\cline{1-10} 
\multirow{2}{*}{jersey\_storm} & \multirow{2}{*}{16.29} & Affected & 382 & 9,088 & 278 & 72.77 & 3.06 & 2.15 & 51.19 \tn 
\multicolumn{1}{| c |}{} & \multicolumn{1}{c |}{}  & Unaffected & 1,307 & 38,093 & 862 & 65.95 & 2.26 & 1.67 & 48.56 \tn\cline{1-10} 
\multirow{2}{*}{california\_fire} & \multirow{2}{*}{17.02} & Affected & 107 & 656 & 71 & 66.36 & 10.82 & 7.16 & 43.88 \tn 
\multicolumn{1}{| c |}{} & \multicolumn{1}{c |}{} & Unaffected & 643 & 130,494 & 233 & 36.24 & 0.18 & 0.18 & 36.71 \tn\cline{1-10} 
\multirow{2}{*}{washington\_mudslide} & \multirow{2}{*}{14.75} & Affected & 174 & 2,774 & 104 & 59.77 & 3.75 & 2.78 & 44.31 \tn 
\multicolumn{1}{| c |}{}  & \multicolumn{1}{c |}{}  & Unaffected & 1,707 & 26,193 & 860 & 50.38 & 3.28 & 2.75 & 42.18 \tn\cline{1-10} 
\hline
\end{tabular}
\caption{Summary of results.}
\label{tab:results}
\end{center}
\end{table*}

\begin{figure}[ht]
	\centering
	\begin{minipage}[b]{0.49\linewidth}
		\includegraphics[width=1\textwidth]{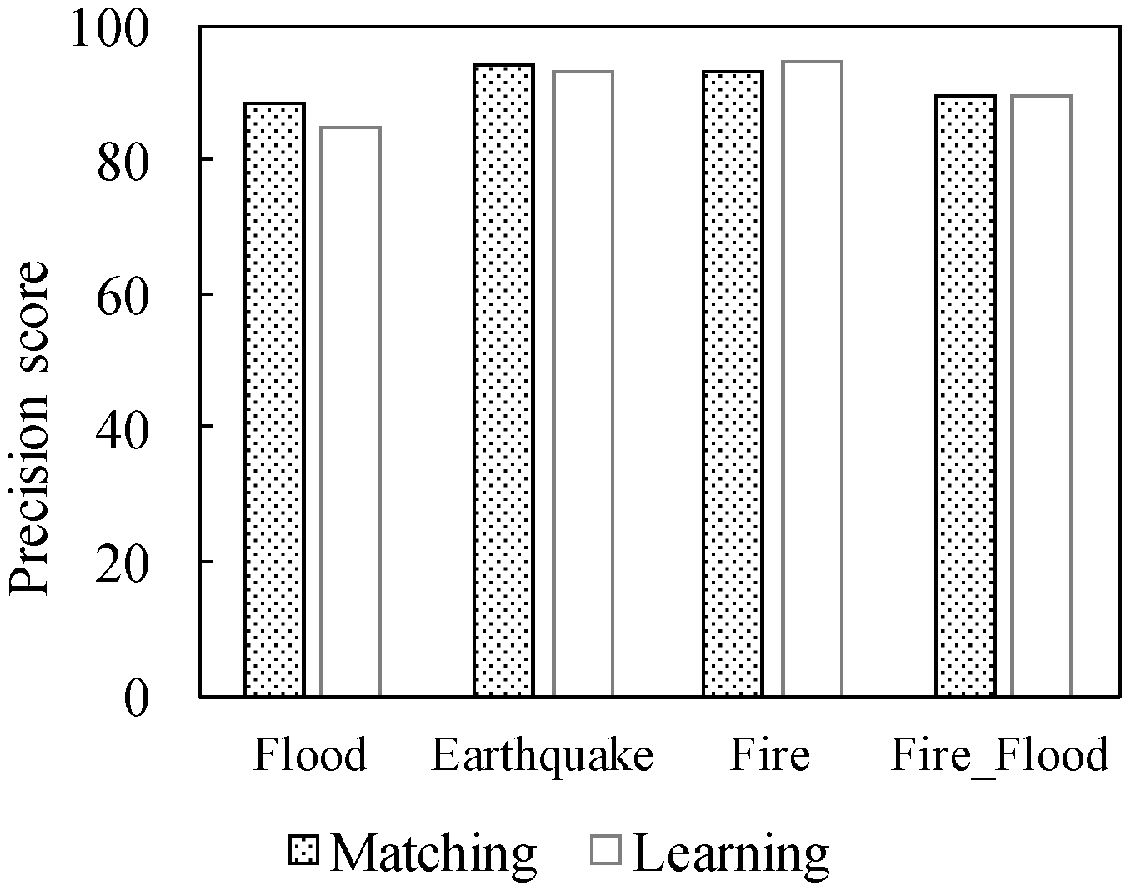}
		\subcaption{Precision score}
		\label{fig:precision-score}
	\end{minipage}
	\centering
	\begin{minipage}[b]{0.49\linewidth}
		\includegraphics[width=1\textwidth]{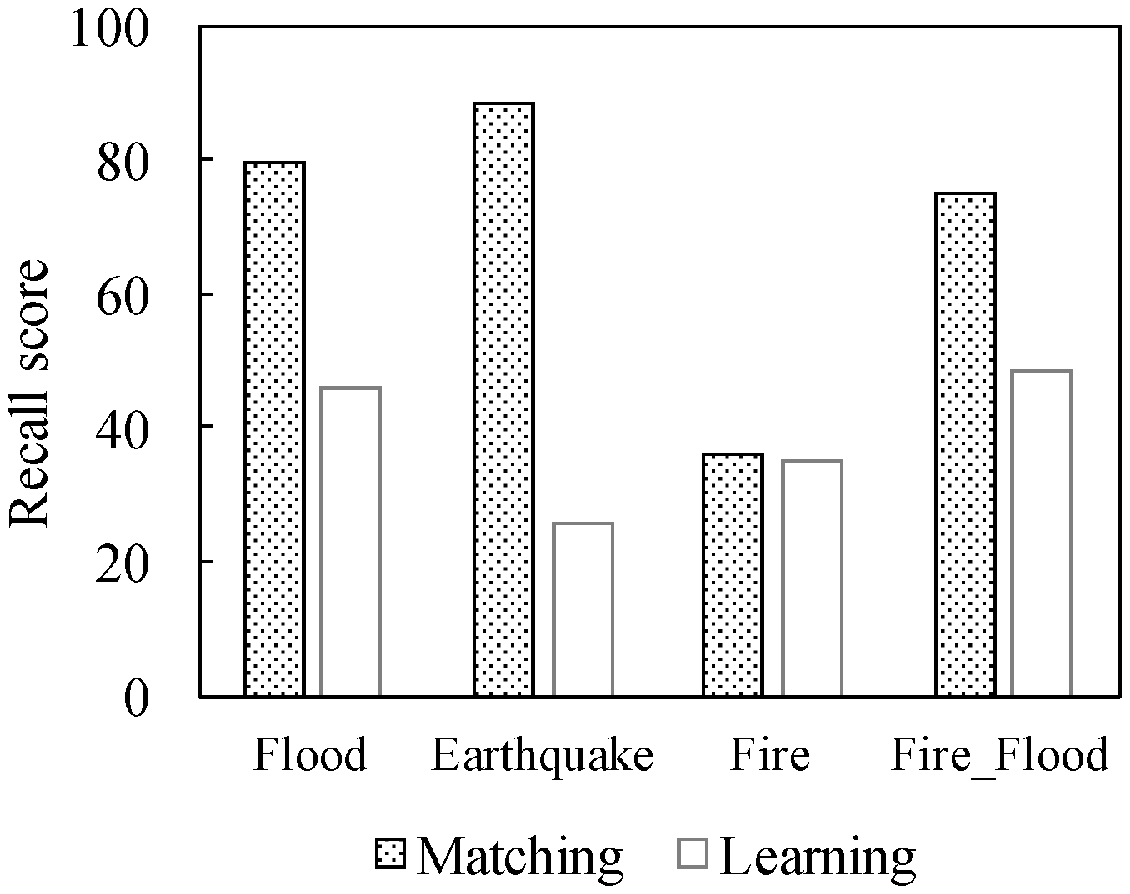}
		\subcaption{Recall score}
		\label{fig:recall-score}
	\end{minipage}
	\caption{Comparison of matching vs. learning-based approaches on labeled dataset.}
	\label{fig:ground_truth}
\end{figure}

Thus far, we assumed that the set of agreed tweets is the ground truth, which may not be true. Hence, we evaluate the matching-based and learning-based approaches in terms of precision and recall scores using the CrisisLexT26 dataset, which has ground truth labeled by the human (see Table~\ref{tab:labeled_datasets}). We split CrisisLexT26 into two equal parts for training and testing. Noting that the matching-based approach does not have the concept of training data; thus, we create hashtags from the entire dataset. The results on the testing data are shown in Figure~\ref{fig:ground_truth}. 
As predicted, two approaches provide high precision scores, between 85\% and 95\% (Figure~\ref{fig:precision-score}). Importantly, when compared to the learning-based approach, the matching-based approach yields much higher recall scores  (Figure~\ref{fig:recall-score}), e.g., 248\% in the earthquake case. The reason for this is that the learning-based approach would include tweets that contain no relevant keywords or hashtags, which may reduce recall.

\subsubsection{The impact of identifying relevant tweets to sentiment analysis application}

\begin{figure}[ht]
	\centering
	\begin{minipage}[b]{0.49\linewidth}
		\includegraphics[width=1\textwidth]{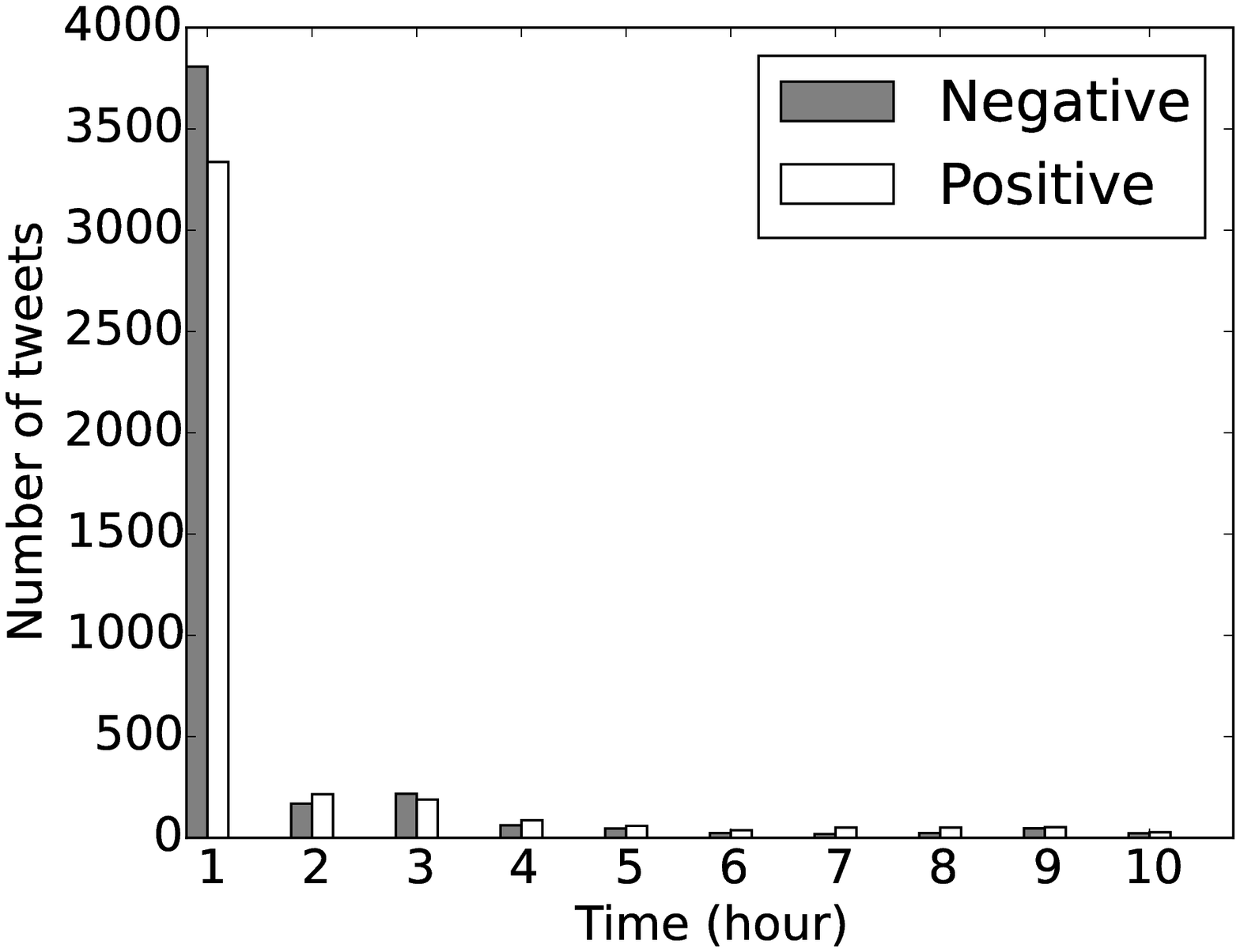}
		\subcaption{Matching (napa\_earthquake)}
		\label{fig:napa_senti_hour}
	\end{minipage}
	\centering
	\begin{minipage}[b]{0.49\linewidth}
		\includegraphics[width=1\textwidth]{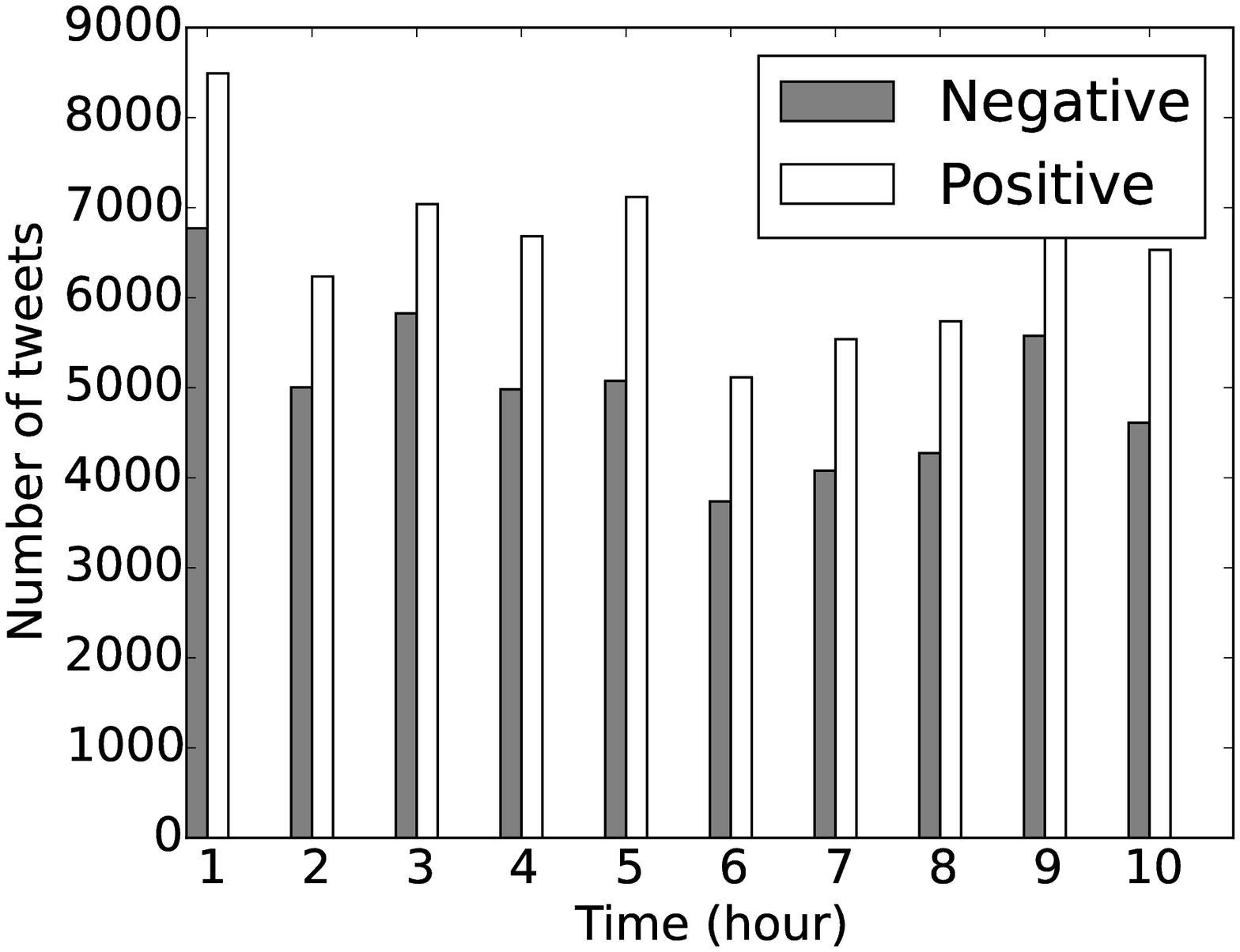}
		\subcaption{Learning (napa\_earthquake)}
		\label{fig:napa_senti_hour_learn}
	\end{minipage}
	\centering
	\begin{minipage}[b]{0.49\linewidth}
		\includegraphics[width=1\textwidth]{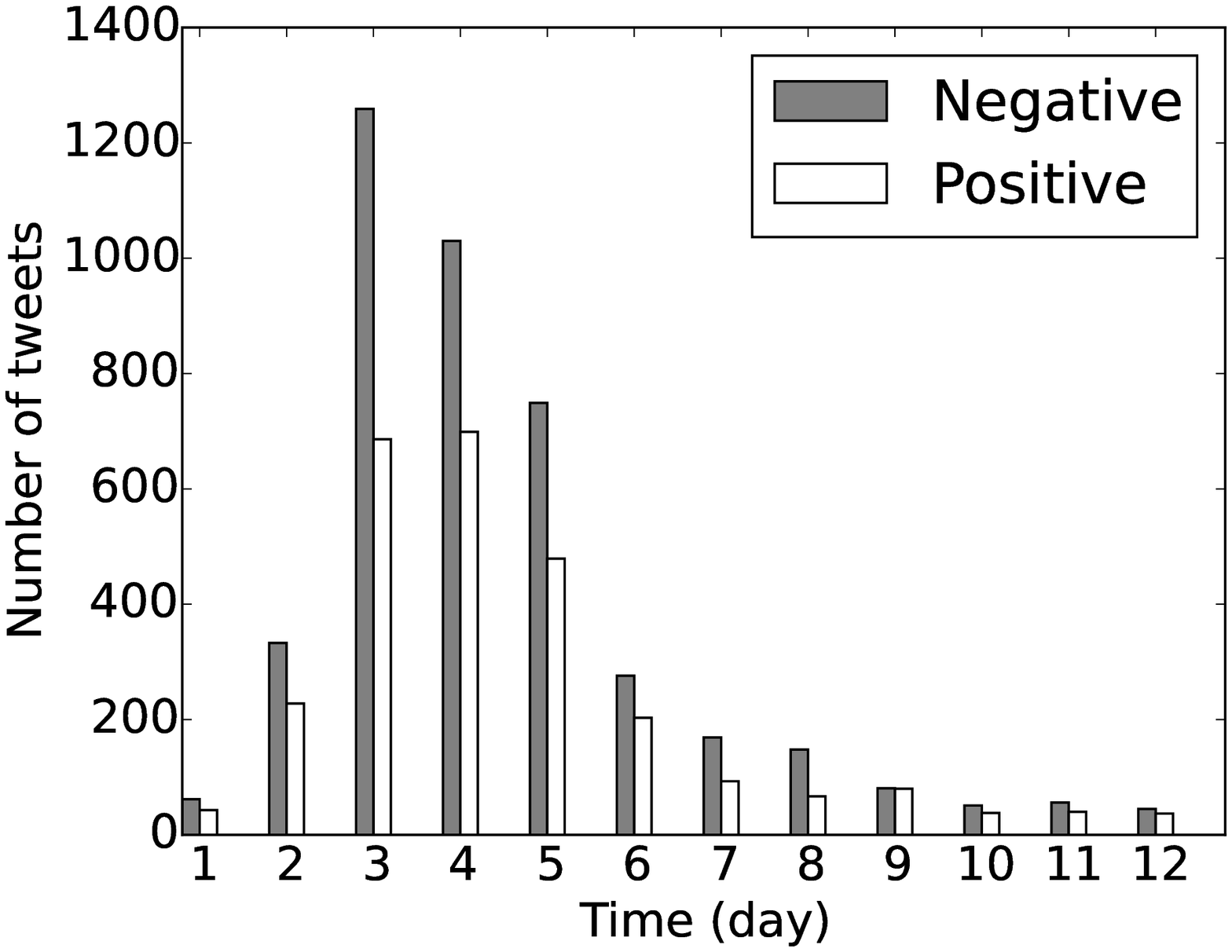}
		\subcaption{Matching (newyork\_storm)}
		\label{fig:ny_senti_day}
	\end{minipage}
	\centering
	\begin{minipage}[b]{0.49\linewidth}
		\includegraphics[width=1\textwidth]{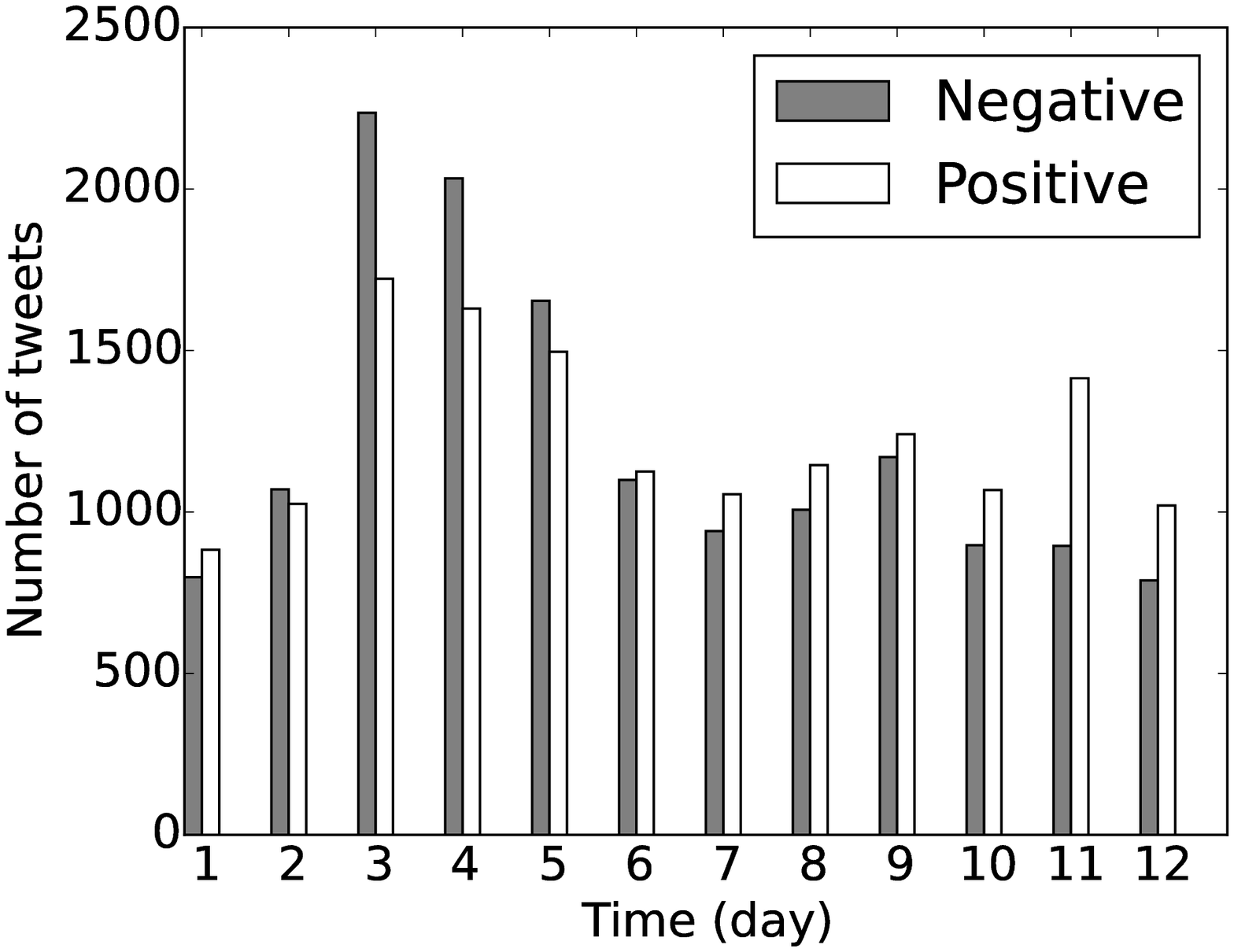}
		\subcaption{Learning (newyork\_storm)}
		\label{fig:ny_senti_day_learn}
	\end{minipage}
	\caption{Impact of matching vs. learning-based approaches to sentiment analysis.}
	\label{fig:sentiment_visualization}
\end{figure}

In this section, we show the importance of accurately obtaining relevant tweets in a popular application of sentiment analysis. Figure~\ref{fig:sentiment_visualization} presents the results of two datasets: the Napa earthquake and the New York Storm. Each figure shows the number of positive and negative tweets in affected area over time (hourly or daily). Figures~\ref{fig:napa_senti_hour} and~\ref{fig:ny_senti_day} show that, during the peak time periods of the disasters (the first hour in earthquake case and days 3,4,5 in storm case), far more negative tweets than positive tweets were posted when people suffered the most.  
Then, the number of relevant tweets diminishes quickly over time, especially in the case of the Napa earthquake when the duration of disaster was short. However, this trend is not clearly shown in Figure~\ref{fig:napa_senti_hour_learn} and~\ref{fig:ny_senti_day_learn} where the learning-based approach was used. This is because the learning approach adds excessive irrelevant tweets that may have higher positive ratio than disaster-related tweets. This result confirms our prior finding that the matching-based technique is better at selecting relevant tweets.

Since the matching-based approach provides the sets of disaster-related tweets of higher quality, we will use this technique from now on.

\subsubsection{Other results}

\paragraph{The impact of extending the set of hashtags}

\begin{figure}[ht]
	\centering
		\includegraphics[width=0.35\textwidth]{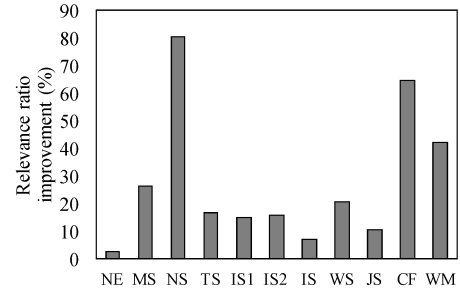}
	\caption{The improvement of our matching-based technique over the conventional method. NS stands for newyork\_storm, etc.}
	\label{fig:matching_improvement}
\end{figure}

In this section, we compare our matching-based approach and the conventional matching-based approach. Figure~\ref{fig:matching_improvement} shows the improvement of our technique over the conventional method in terms of selecting the number of relevant tweets. We observe that our approach produced a significantly bigger dataset of relevant tweets when compared to conventional approaches. The reason for the increase is that our matching-based technique is able to include a more diverse set of hashtags (e.g., \emph{\#3amearthquake, \#quakeinsf}). This improvement in terms of quality is even more important given the high recall score of our matching-based approach. We also observe that the improvement is higher in populated areas such as New York because there are more unique hashtags being used. 

\paragraph{The impact of removing spam tweets on the number of relevant tweets}

\begin{figure}[ht]
	\centering
	\begin{minipage}[b]{0.49\linewidth}
		\includegraphics[width=1\textwidth]{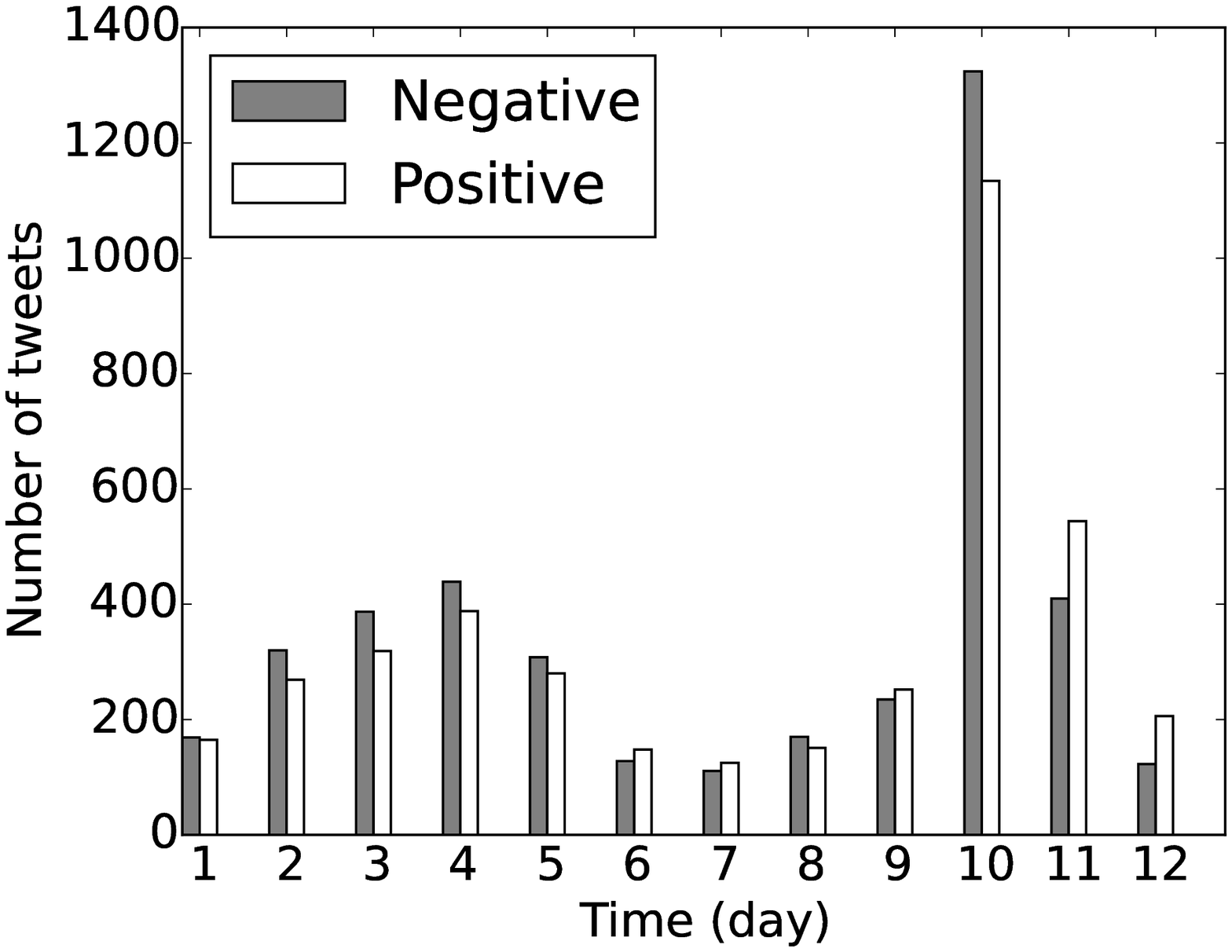}
		\subcaption{With spam tweets}
		\label{fig:ny_senti_day_unaffected2}
	\end{minipage}
	\centering
	\begin{minipage}[b]{0.49\linewidth}
		\includegraphics[width=1\textwidth]{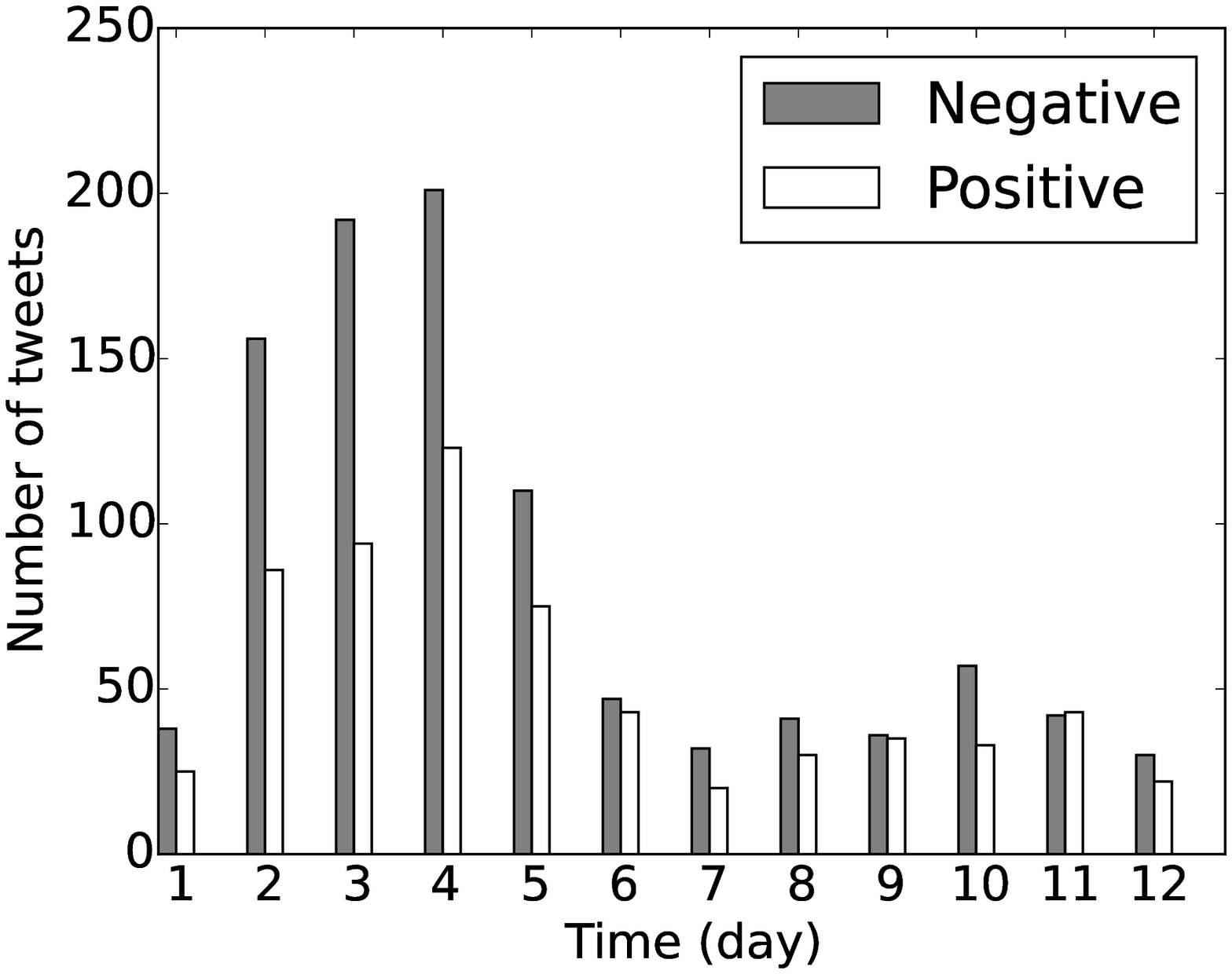}
		\subcaption{Without spam tweets}
		\label{fig:ny_senti_day_unaffected}
	\end{minipage}
	\caption{Impact of removing spam tweets to the number of tweets on unaffected region of the newyork\_storm dataset.}
	\label{fig:spam_sentiment}
\end{figure}

Table~\ref{tab:results} shows the percentage of spam tweets for each disaster, referred to as \emph{spam ratio}. The spam ratio ranges from 14\% to 29\%. In removing the spam tweets, we observe that although the percentage of the spam users is only 0.8\% (1,937 spammers over 144,297 unique users), they generate 23.33\% of the total tweets (928,174 spam tweets over 3,978,713 total tweets). These statistics show that eliminating spam tweets is an important step in real applications.
To illustrate, Figure~\ref{fig:spam_sentiment} shows the impact of removing spam tweets on the number of positive/negative tweets obtained. Figure~\ref{fig:ny_senti_day_unaffected2} shows an unexpected peak in day 10, which disappears after removing spam tweets (see Figure~\ref{fig:ny_senti_day_unaffected}). We further identify a set of 50 active spammers, each posted a large number of tweets (331 to 1121) on that day; most tweets from a spammer have the same content.


\section{Conclusion and Future Works}
\label{sec:conclude}

In this paper we introduced a five-step framework for analyzing tweets during disasters to enhance situational awareness. We proposed two different techniques for identifying tweets relevant to a particular disaster type. We observed that the matching-based technique includes less relevant tweets but with higher quality when compared to the learning-based approach. We confirmed our finding by conducting various experiments on three types of disasters (earthquake, flood, wildfire).
As a future work, we aim to further evaluate trade-offs between the two techniques by obtaining the ground truth of the datasets through crowdsourcing using Amazon Mechanical Turk. We will also extend our framework to include man-made disasters such as terrorism. Another challenging problem to address is real-time classification of disaster-related tweets for timely analysis of the data. 
Finally, we aim to use the proposed framework to construct sentiment maps of the affected areas for real-time situational awareness during disasters. This is a step toward using the classified and sentimented information in real-world situation, which is often overlooked in the existing studies. 



\section*{Acknowledgement}
This research has been funded by NSF grants IIS-1320149, CNS-1461963 and the USC Integrated Media Systems Center. Any opinions, findings, and conclusions or recommendations expressed in this material are those of the authors and do not necessarily reflect the views of any of the sponsors such as NSF.

\bibliographystyle{abbrv}
\bibliography{sigproc}

\begin{thebibliography}{10}

\bibitem{acar2011twitter}
A.~Acar and Y.~Muraki.
\newblock Twitter for crisis communication: lessons learned from japan's
  tsunami disaster.
\newblock {\em International Journal of Web Based Communities}, 7(3):392--402,
  2011.

\bibitem{ashktorab2014tweedr}
Z.~Ashktorab, C.~Brown, M.~Nandi, and A.~Culotta.
\newblock Tweedr: Mining twitter to inform disaster response.
\newblock {\em Proc. of ISCRAM}, 2014.

\bibitem{caragea2011classifying}
C.~Caragea, N.~McNeese, A.~Jaiswal, G.~Traylor, H.-W. Kim, P.~Mitra, D.~Wu,
  A.~H. Tapia, L.~Giles, B.~J. Jansen, et~al.
\newblock Classifying text messages for the haiti earthquake.
\newblock In {\em Proceedings of the 8th international conference on
  information systems for crisis response and management (ISCRAM2011)}.
  Citeseer, 2011.

\bibitem{caragea2016identifying}
C.~Caragea, A.~Silvescu, and A.~H. Tapia.
\newblock Identifying informative messages in disaster events using
  convolutional neural networks.
\newblock In {\em International Conference on Information Systems for Crisis
  Response and Management}, 2016.

\bibitem{caragea2014mapping}
C.~Caragea, A.~Squicciarini, S.~Stehle, K.~Neppalli, and A.~Tapia.
\newblock Mapping moods: geo-mapped sentiment analysis during hurricane sandy.
\newblock {\em Proc. of ISCRAM}, 2014.

\bibitem{cobo2015identifying}
A.~Cobo, D.~Parra, and J.~Nav{\'o}n.
\newblock Identifying relevant messages in a twitter-based citizen channel for
  natural disaster situations.
\newblock In {\em Proceedings of the 24th International Conference on World
  Wide Web}, pages 1189--1194. ACM, 2015.

\bibitem{de2012unfolding}
M.~De~Choudhury, N.~Diakopoulos, and M.~Naaman.
\newblock Unfolding the event landscape on twitter: classification and
  exploration of user categories.
\newblock In {\em Proceedings of the ACM 2012 conference on Computer Supported
  Cooperative Work}, pages 241--244. ACM, 2012.

\bibitem{diakopoulos2012finding}
N.~Diakopoulos, M.~De~Choudhury, and M.~Naaman.
\newblock Finding and assessing social media information sources in the context
  of journalism.
\newblock In {\em Proceedings of the SIGCHI Conference on Human Factors in
  Computing Systems}, pages 2451--2460. ACM, 2012.

\bibitem{fema2012crisis}
FEMA.
\newblock Situational awareness in mass emergency: A behavioral and linguistic
  analysis of microblogged communications.
\newblock 2012.

\bibitem{fraustino2012social}
J.~D. Fraustino, B.~Liu, and Y.~Jin.
\newblock Social media use during disasters: a review of the knowledge base and
  gaps.
\newblock 2012.

\bibitem{hu2013unsupervised}
X.~Hu, J.~Tang, H.~Gao, and H.~Liu.
\newblock Unsupervised sentiment analysis with emotional signals.
\newblock In {\em Proceedings of the 22nd international conference on World
  Wide Web}, pages 607--618. ACM, 2013.

\bibitem{hu2013exploiting}
X.~Hu, L.~Tang, J.~Tang, and H.~Liu.
\newblock Exploiting social relations for sentiment analysis in microblogging.
\newblock In {\em Proceedings of the sixth ACM international conference on Web
  search and data mining}, pages 537--546. ACM, 2013.

\bibitem{imran2015processing}
M.~Imran, C.~Castillo, F.~Diaz, and S.~Vieweg.
\newblock Processing social media messages in mass emergency: A survey.
\newblock {\em ACM Computing Surveys (CSUR)}, 47(4):67, 2015.

\bibitem{imran2014aidr}
M.~Imran, C.~Castillo, J.~Lucas, P.~Meier, and S.~Vieweg.
\newblock {AIDR}: Artificial intelligence for disaster response.
\newblock In {\em Proceedings of the 23rd International Conference on World
  Wide Web}, pages 159--162. ACM, 2014.

\bibitem{imran2013practical}
M.~Imran, S.~Elbassuoni, C.~Castillo, F.~Diaz, and P.~Meier.
\newblock Practical extraction of disaster-relevant information from social
  media.
\newblock In {\em Proceedings of the 22nd International Conference on World
  Wide Web}, pages 1021--1024. ACM, 2013.

\bibitem{imran2013extracting}
M.~Imran, S.~M. Elbassuoni, C.~Castillo, F.~Diaz, and P.~Meier.
\newblock Extracting information nuggets from disaster-related messages in
  social media.
\newblock {\em Proc. of ISCRAM, Baden-Baden, Germany}, 2013.

\bibitem{kanhabua2013understanding}
N.~Kanhabua and W.~Nejdl.
\newblock Understanding the diversity of tweets in the time of outbreaks.
\newblock In {\em Proceedings of the 22nd International Conference on World
  Wide Web}, pages 1335--1342. ACM, 2013.

\bibitem{le2014distributed}
Q.~V. Le and T.~Mikolov.
\newblock Distributed representations of sentences and documents.
\newblock In {\em ICML}, volume~14, pages 1188--1196, 2014.

\bibitem{liu2012sentiment}
B.~Liu.
\newblock Sentiment analysis and opinion mining.
\newblock {\em Synthesis lectures on human language technologies}, 5(1):1--167,
  2012.

\bibitem{lu2015visualizing}
Y.~Lu, X.~Hu, F.~Wang, S.~Kumar, H.~Liu, and R.~Maciejewski.
\newblock Visualizing social media sentiment in disaster scenarios.
\newblock In {\em Proceedings of the 24th International Conference on World
  Wide Web}, pages 1211--1215. ACM, 2015.

\bibitem{mikolov2013distributed}
T.~Mikolov, I.~Sutskever, K.~Chen, G.~S. Corrado, and J.~Dean.
\newblock Distributed representations of words and phrases and their
  compositionality.
\newblock In {\em Advances in neural information processing systems}, pages
  3111--3119, 2013.

\bibitem{monroy2013new}
A.~Monroy-Hern{\'a}ndez, E.~Kiciman, M.~De~Choudhury, S.~Counts, et~al.
\newblock The new war correspondents: The rise of civic media curation in urban
  warfare.
\newblock In {\em Proceedings of the 2013 conference on Computer supported
  cooperative work}, pages 1443--1452. ACM, 2013.

\bibitem{morstatter2013sample}
F.~Morstatter, J.~Pfeffer, H.~Liu, and K.~M. Carley.
\newblock Is the sample good enough? comparing data from twitter's streaming
  api with twitter's firehose.
\newblock {\em arXiv preprint arXiv:1306.5204}, 2013.

\bibitem{nguyen2016applications}
D.~T. Nguyen, S.~Joty, M.~Imran, H.~Sajjad, and P.~Mitra.
\newblock Applications of online deep learning for crisis response using social
  media information.
\newblock {\em arXiv preprint arXiv:1610.01030}, 2016.

\bibitem{olteanu2014crisislex}
A.~Olteanu, C.~Castillo, F.~Diaz, and S.~Vieweg.
\newblock Crisislex: A lexicon for collecting and filtering microblogged
  communications in crises.
\newblock In {\em ICWSM}, 2014.

\bibitem{olteanu2015expect}
A.~Olteanu, S.~Vieweg, and C.~Castillo.
\newblock What to expect when the unexpected happens: Social media
  communications across crises.
\newblock In {\em Proceedings of the 18th ACM Conference on Computer Supported
  Cooperative Work \& Social Computing}, pages 994--1009. ACM, 2015.

\bibitem{pang2008opinion}
B.~Pang and L.~Lee.
\newblock Opinion mining and sentiment analysis.
\newblock {\em Foundations and trends in information retrieval}, 2(1-2):1--135,
  2008.

\bibitem{parilla2014disaster}
B.~E. Parilla-Ferrer, P.~L. Fernandez, and J.~T. Ballena.
\newblock {Automatic Classification of Disaster-Related Tweets}.
\newblock In {\em Proc. International conference on Innovative Engineering
  Technologies ({ICIET})}, pages 62+, Dec. 2014.

\bibitem{pfeffer2016geotagged}
J.~Pfeffer and F.~Morstatter.
\newblock Geotagged twitter posts from the united states: A tweet collection to
  investigate representativeness.
\newblock {\em Version: 1. GESIS Data Archive. Dataset.
  http://doi.org/10.7802/1166}, 2016.

\bibitem{rehurek_lrec}
R.~{\v R}eh{\r u}{\v r}ek and P.~Sojka.
\newblock {Software Framework for Topic Modelling with Large Corpora}.
\newblock In {\em {Proceedings of the LREC 2010 Workshop on New Challenges for
  NLP Frameworks}}, pages 45--50, Valletta, Malta, May 2010. ELRA.

\bibitem{sakaki2010earthquake}
T.~Sakaki, M.~Okazaki, and Y.~Matsuo.
\newblock Earthquake shakes twitter users: real-time event detection by social
  sensors.
\newblock In {\em Proceedings of the 19th international conference on World
  wide web}, pages 851--860. ACM, 2010.

\bibitem{starbird2011voluntweeters}
K.~Starbird and L.~Palen.
\newblock Voluntweeters: Self-organizing by digital volunteers in times of
  crisis.
\newblock In {\em Proceedings of the SIGCHI Conference on Human Factors in
  Computing Systems}, pages 1071--1080. ACM, 2011.

\bibitem{vieweg2010microblogging}
S.~Vieweg, A.~L. Hughes, K.~Starbird, and L.~Palen.
\newblock Microblogging during two natural hazards events: what twitter may
  contribute to situational awareness.
\newblock In {\em Proceedings of the SIGCHI conference on human factors in
  computing systems}, pages 1079--1088. ACM, 2010.

\bibitem{vieweg2012situational}
S.~E. Vieweg.
\newblock Situational awareness in mass emergency: A behavioral and linguistic
  analysis of microblogged communications.
\newblock 2012.

\bibitem{vu2016geosocialbound}
D.~D. Vu, H.~To, W.-Y. Shin, and C.~Shahabi.
\newblock Geosocialbound: an efficient framework for estimating social poi
  boundaries using spatio--textual information.
\newblock In {\em Proceedings of the Third International ACM SIGMOD Workshop on
  Managing and Mining Enriched Geo-Spatial Data}, page~3. ACM, 2016.

\bibitem{zhang2016semi}
S.~Zhang and S.~Vucetic.
\newblock Semi-supervised discovery of informative tweets during the emerging
  disasters.
\newblock {\em arXiv preprint arXiv:1610.03750}, 2016.

\end{thebibliography}

\balance
\end{document}